\documentclass[showpacs,amsmath,amssymb,aps,10pt,reprint,superscriptaddress]{revtex4-1}
\usepackage{graphicx}
\usepackage{bm}
\usepackage[breaklinks=true,colorlinks=true,linkcolor=blue,urlcolor=blue,citecolor=blue]{hyperref}
\usepackage{graphics}
\usepackage{dcolumn}
\usepackage{amsmath,amssymb}
\usepackage{mathdots}
\usepackage{natbib}
\usepackage{soul,color}
\usepackage{epstopdf}
\usepackage[note-name]{notes2bib}

\begin{document}
\title{\Large Spin-wave phase inverter upon a single nanodefect}

\author{O. V.~Dobrovolskiy}
    \affiliation{Physikalisches Institut, Goethe University, 60438 Frankfurt am Main, Germany}
    \affiliation{Physics Department, V. N. Karazin Kharkiv National University, 61022 Kharkiv, Ukraine}
\author{R. Sachser}
    \affiliation{Physikalisches Institut, Goethe University, 60438 Frankfurt am Main, Germany}
\author{S. A.~Bunyaev}
\author{D.~Navas}
    \affiliation{Dpto de Fisica e Astronomia, University of Porto IFIMUP-IN, 4169-007 Porto, Portugal}
\author{\\V.~M.~Bevz}
    \affiliation{Physics Department, V. N. Karazin Kharkiv National University, 61022 Kharkiv, Ukraine}
    \affiliation{ICST Faculty, Ukrainian State University of Railway Transport, 61050 Kharkiv, Ukraine}
\author{M. Zelent}
    \affiliation{Faculty of Physics, Adam Mickiewicz University, Poznan, Poland}
\author{W. \'Smigaj}
    \affiliation{Synopsys Ltd., Bradninch Hall, Castle Street, EX4 3PL, Exeter}
\author{J.~Rychly}
\author{M.~Krawczyk}
    \affiliation{Faculty of Physics, Adam Mickiewicz University, Poznan, Poland}
\author{\\R.~V.~Vovk}
    \affiliation{Physics Department, V. N. Karazin Kharkiv National University, 61022 Kharkiv, Ukraine}
    \affiliation{ICST Faculty, Ukrainian State University of Railway Transport, 61050 Kharkiv, Ukraine}
\author{M. Huth}
    \affiliation{Physikalisches Institut, Goethe University, 60438 Frankfurt am Main, Germany}
\author{G.~N.~Kakazei}
    \affiliation{Dpto de Fisica e Astronomia, University of Porto IFIMUP-IN, 4169-007 Porto, Portugal}

\date{\today}

\begin{abstract}
Local modification of magnetic properties of nanoelements is a key to design future-generation magnonic devices, in which information is carried and processed via spin waves. One of the biggest challenges here is to fabricate simple and miniature phase-controlling elements with broad tunability. Here, we successfully realize such spin-wave phase shifter upon a single nanogroove milled by focused ion beam in a Co-Fe microsized magnonic waveguide. By varying the groove depth and the in-plane bias magnetic field we continuously tune the spin-wave phase and experimentally evidence a complete phase inversion. The microscopic mechanism of the phase shift is based on the combined action of the nanogroove as a geometrical defect and the lower spin-wave group velocity in the waveguide under the groove where the magnetization is reduced due to the incorporation of Ga ions during the ion-beam milling. The proposed phase shifter can easily be on-chip integrated with spin-wave logic gates and other magnonic devices. Our findings are crucial for designing nano-magnonic circuits and for the development of spin-wave nano-optics.
\end{abstract}
\maketitle


\section{Introduction}
Spin waves, and their quanta magnons, can be used for carrying and processing of information without net transfer of electric charge \cite{Khi10jpd,Kru10jpd,Chu17jpd}. Featuring low dissipation, magnonic devices based on spin-wave logic gates are being actively developed to expand the functionality of next-generation nanoelectronics and information processing. For instance, prototypes of a spin wave-based NOT gate \cite{Kos05apl}, a majority gate \cite{Fis17apl} as well as exclusive-not-OR and not-AND gates based on a Mach-Zehnder-type interferometer \cite{Sch08apl} have been demonstrated experimentally. In these, logic information is encoded in the phase of spin waves, such that logic ``0'' corresponds to a certain phase $\phi^{(0)}$ while logic ``1'' is represented by the phase of $\phi^{(1)} = \phi^{(0)} + \pi$ \cite{Fis17apl}. Spin-wave phase-setting elements are therefore important for magnonic information processing. Furthermore, the interest in tailoring spin-wave phase in magnonics is urged by the implementation of concepts originating from optics. Here, the aim is to achieve steering of spin waves in conjunction with the miniaturization of a device while maintaining its high-frequency operation \cite{Chu15nph}. The emerging domain of spin-wave nano-optics \cite{Dem08apl} requires the generation and manipulation of spin wave beams in the microwave frequency range, thus enabling realization of spin-wave fibers \cite{Yuw16prb} and exploitation of refraction and reflection effects \cite{Gru14apl,Sti16prl}.

Previously, it has been shown that a thickness step can be used as an interface between two media with different dispersion relations for tailoring the spin-wave refraction \cite{Sti16prl}. Recently, it has been predicted \cite{Gru17prb} and experimentally demonstrated \cite{Sti18prl} that the spin-wave beam spot can be shifted along the interface due to the Goos-H\"anchen effect stipulated by the reflections within the finite-sized beam interfering along the line transverse to the average propagation direction. In addition, the propagation of spin waves through areas with changed magnetization is also widely exploited in magnonic crystals \cite{Kra14pcm,Liu13apl}. In such crystals, which are media with periodically varying magnetic characteristics, spin waves exhibit rejection bands where the spin-wave propagation is prohibited. Under proper conditions, individual building blocks of magnonic crystals allow for control of spin-wave amplitude and phase \cite{Hub13prb}. In particular, it was shown both experimentally and by numerical simulations that spin waves propagating in a magnetic film can pass through a region of a magnetic field inhomogeneity \cite{Dem04prl} or a mechanical gap \cite{Sch10epl}. Furthermore, spin waves were predicted to change their phase when passing through a magnetic domain wall in different branches of a ring \cite{Her04prl} or when a local spin-polarized current is applied to a ferromagnetic stripe \cite{Che15mmm}. Due to experimental challenges, these approaches have not been realized so far. Further systems suggested as spin-wave phase shifters include magnetic dot arrays with a chessboard antiferromagnetic ground state \cite{Lou16aip}, multiferroic structures \cite{Ust14apl}, application of polarized light pulses \cite{Yos14jap}, electric fields \cite{Wan18jap}, nanomagnets \cite{Auy12apl}, and increasing signal power \cite{Han09apl}. While the lateral dimensions of the key elements of the experimentally realized devices are in the millimeter \cite{Ust14apl,Yos14jap} or $10\,\mu$m-range \cite{Han09apl,Bau18apl}, few studies were concerned with phase shifts of spin waves as they propagate through a nano-sized defect. Remarkably, phase shifts of up to almost $\pi$ have been observed for spin waves in a one-dimensional magnonic crystal \cite{Bau18apl}. There, the phase accumulation was explained by a modified wave vector $k$ at a single $8\,\mu$m-long magnetic defect acting as a spin-wave well \cite{Bau18apl}. Thus, among current challenges in nano-magnonics are downscaling of phase-setting elements in the nanometer range while keeping them simple, programmable, and integrable with other circuit elements. For instance, the ability to fine-tune the spin-wave phase front at the lateral nanoscale would enable the fabrication of spin-wave lenses acting by analogy with phased-array antennas for electromagnetic waves. In return, this would allow for an essential supplement of the existing approaches to generate and steer spin-wave beams by tailoring the geometry of electromagnetic-to-spin wave transducers \cite{Gru16nsr,Kor17prb}.

Here, we investigate both experimentally and by micromagnetic simulations the evolution of phase of spin waves propagating through $1\,\mu$m-wide Co-Fe waveguides containing a single nanogroove milled by focused ion beam (FIB). By increasing the groove depth and the in-plane biasing magnetic field we continuously tune the spin-wave phase and experimentally evidence a complete phase inversion. Mapping of the experimental results onto micromagnetic simulations reveals that for the complete phase inversion it is not enough to treat the nanogroove just as a geometric thickness reduction. Instead, the microscopic mechanism of phase inversion relies largely on the reduced magnetization under the groove where the phase velocity of the spin waves differs essentially from that in the plain waveguide. In all, our findings are crucial for phase-tuning of spin waves in magnonic circuits, further development of spin-wave logic devices, and pave the way to the fabrication of tunable magnonic lenses in spin-wave nano-optics.

\section{Results and discussion}
\begin{figure}
    \centering
    \includegraphics[width=1\linewidth]{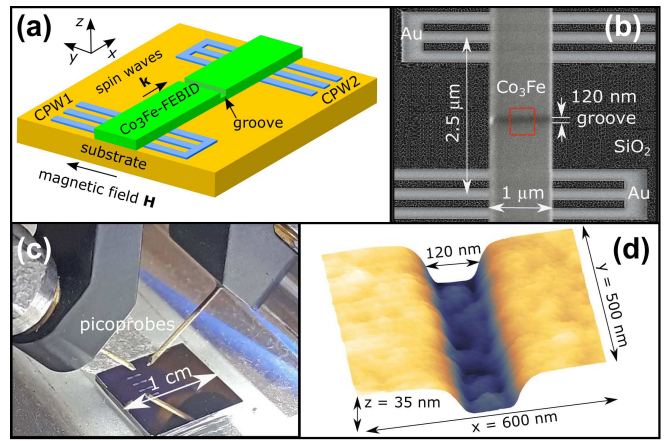}
    \caption{\textbf{Investigated system}. (a) Experimental geometry. A Co-Fe magnonic waveguide is fabricated by focused electron beam-induced deposition, bridging two microwave antennas (CPW1 and CPW2) which are used for the excitation and detection of transmitted spin waves. A nanogroove milled by focused ion beam in the middle of the waveguide causes a phase shift in spin waves passing through it. An in-plane biasing magnetic field $\mathbf{H}$ is applied perpendicular to the direction of propagation of spin waves and parallel to the nanogroove. The oxide layer on top of the CPWs, which electrically isolates the magnonic waveguide from the CPWs, is not shown for simplicity. (b) Scanning electron microscopy image of the waveguide with a 35\,nm-deep nanogroove. The dashed rectangle encages the region corresponding to the atomic microscopy image of the sample surface shown in (d). (c) The CPWs are connected to the vector network analyzer via two picoprobes.}
    \label{fGeometry}
\end{figure}

The investigated system and the experimental geometry are shown in Fig. \ref{fGeometry}. The device is realized upon a $1\,\mu$m-wide Co-Fe waveguide with thickness $t = 45$\,nm fabricated by focused electron beam-induced deposition (FEBID). A nanogroove milled by FIB in the middle of the waveguide causes a phase shift in spin waves passing through it. The half-depth width of the groove $b$ is equal to $120$\,nm. An in-plane biasing magnetic field $\mathbf{H}$ is applied perpendicular to the direction of propagation of spin waves and parallel to the nanogroove, thus establishing the conditions for the propagation of spin waves in the Damon-Eshbach geometry. Two microwave antennas, CPW1 and CPW2, are used for the excitation and detection of transmitted spin waves by all-electrical spin-wave spectroscopy \cite{Neu10prl,Tac12prb} employing a vector network analyzer. Transmission of spin waves in four waveguides with the groove depth $d$ varied between $5$\,nm and $35$\,nm in steps of 10\,nm is compared with a reference waveguide without groove. Further details on the fabrication and characterization of the samples are provided in Sec. \ref{sExp}.

\begin{figure}
    \centering
    \includegraphics[width=1\linewidth]{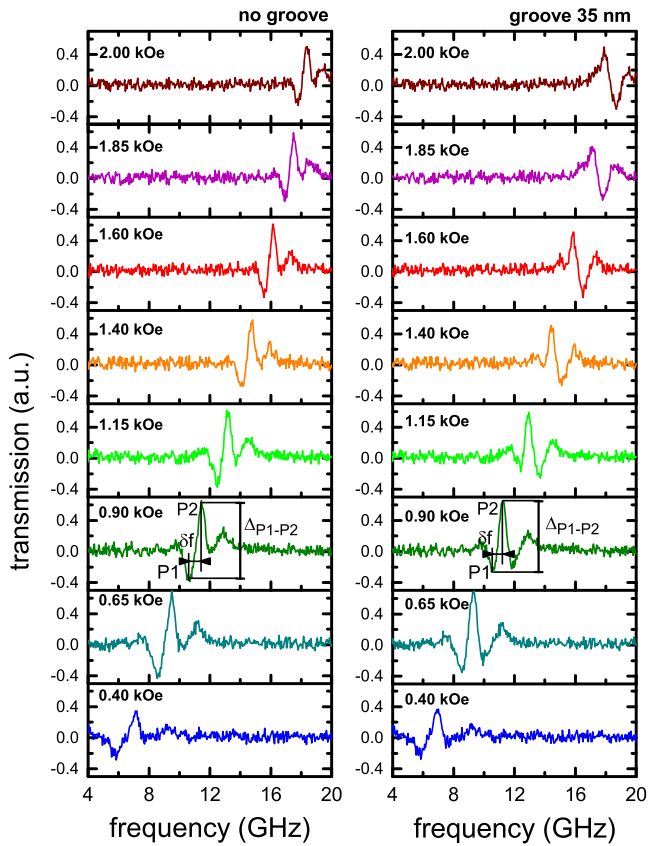}
    \caption{\textbf{Spin-wave transmission through the waveguides}. Frequency traces of the magnitude of $S_{21}(f)$ for a series of magnetic fields $H$, as indicated. Left column: reference waveguide. Right column: waveguide with a 35\,nm-deep nanogroove. The extrema $P1$ and $P2$, the relative signal strength $\Delta_{P2-P1}$, and the frequency difference $\delta f = f_{P2}-f_{P1}$ between them are indicated.}
    \label{fMain}
\end{figure}

\textcolor{blue}{Figure} \ref{fMain} presents the evolution of the spin-wave transmission with increasing magnetic field for the plain sample (left column) in comparison with the sample with the $35$\,nm-deep groove (right column). The transmitted signal oscillates and the envelope's central frequency is increasing with increasing $H$. This is a fingerprint of spin-wave propagation. As is known, the otherwise Lorentzian-shaped spin-wave rf susceptibility splits up in a multipeak structure if spin-wave propagation occurs between emitter and detector CPWs \cite{Bai03apl}. The frequency period $\delta f$ between maxima and minima is a measure of the phase difference $\delta\phi$ which spin waves of different eigenfrequencies acquire along the propagation path $L$. The $\delta f$ thus defined corresponds to the change $\delta \phi = \pi$ of the spin-wave phases and the change $\delta k = \delta \phi /L = \pi /L$ of the spin-wave wave vectors. The corresponding group velocity $v_g$ of spin waves can be extracted from the relation $v_g = \partial\omega /\partial k\approx 2\pi\delta f/(2\pi/L) = \delta f L$ \cite{Neu10prl}. Accordingly, with $L= 2.5\,\mu$m and $2\delta f = 1.2\,$GHz, as deduced for the plain waveguide at $H = 2$\,kOe, we deduce $v_g \approx 3$\,km/s. Further details on the spin-wave transmission and the deduced parameters are provided in \textcolor{red}{Supporting Information}.

\begin{figure}
    \centering
    \includegraphics[width=1\linewidth]{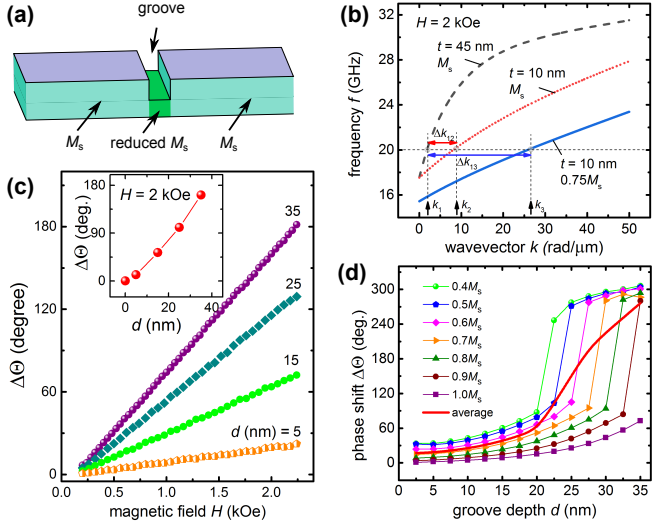}
    \caption{\textbf{Spin-wave phase tuning by magnetic field and groove depth variation}.
    (a) Geometry used in the simulations. The groove is modeled as a rectangular-shaped reduction of the film thickness. The film slice under the groove has a reduced magnetization with respect to $M_\mathrm{s}$ in the as-deposited structure.
    (b) Calculated dispersion relations at $H = 2$\,kOe for a plain film with $M_\mathrm{s} = 1.4 \times 10^6$\,A/m and the thickness $t=45$\,nm (dashed line), $t=10$\,nm (dotted line), and a $10$\,nm-thick film with $0.75M_\mathrm{s}$.
    (c) Experimentally measured phase shift $\Delta\Theta$ as a function of the magnetic field $H$ in the four waveguides with nanogrooves.
    Inset: Experimentally deduced phase shift $\Delta\Theta$ as a function of the groove depth $d$ at $H = 2$\,kOe.
    (d) Calculated phase shift at $20$\,GHz as a function of the nanogroove depth for a $45$\,nm-thick film with the magnetization $M_\mathrm{s}$ in comparison with the cases when a $10$\,-nm thick slice under the groove has a reduced magnetization, as indicated. The solid line is an average of the calculated curves.
}
    \label{fExperiment}
\end{figure}

For the phase-shift deduction, we identify two most pronounced neighboring extrema, $P1$ and $P2$, in both data sets in \textcolor{blue}{Figure} \ref{fMain} and deduce the maximum-to-minimum amplitude $\Delta_{P1-P2}$ (relative signal strength) and the frequency difference $\delta f = f_{P2}-f_{P1}$ between them. To quantify the amplitude change and the frequency and phase shifts in the waveguides with nanogrooves, we introduce $\Delta f = f^{\mathrm{groove}}_{P2} -f^{\mathrm{ref}}_{P2}$ evaluated at different $H$ for the $P2$ peaks for the grooved and reference samples. Since the phase shift between the neighboring oscillation extrema amounts to $\pi$ \cite{Neu10prl,Bau18apl}, we deduce the spin-wave phase shift induced by the nanogroove using the relation $\Delta \Theta = - \pi \Delta f (H)/\delta f(H)$. For all waveguides with a nanogroove, $\Delta \Theta$ is found to increase almost linearly with $H$ and $d$, Fig. \ref{fExperiment}(c). Importantly, a phase shift of about $\pi$ is observed for the 35\,nm-deep groove at $2.25$\,kOe. This means that the whole range of spin wave phases can be achieved for spin waves in this waveguide, while the spin-wave phase can be finely tuned by the biasing magnetic field and coarsely by the groove depth.

Prior to the elucidation of the phase-shift origin in the investigated system, we note that in general one might expect several distinct microscopic mechanisms at work in waveguides with FIB-milled nanogrooves. Firstly, a nanogroove introduces a dynamic demagnetizing field at its edges where spin waves tend to get pinned \cite{Lan17prba} that affects their dynamics. Secondly, one may expect a modified dipolar coupling and a different effective anisotropy constant at the nanogroove edges \cite{Klo12prb}. Finally, during the FIB milling process, implantation of Ga ions is possible at the bottom of the milled nanogroove \cite{Dob12njp}. In return, this may lead to a local modification of the magnetic properties of the waveguide region under the groove, such as magnetization reduction \cite{Dob15bjn}. If one considers the groove solely as a gap under which the waveguide thickness is reduced, to be referred to as a geometric phase-shift mechanism, then a rough estimate for the phase shift is possible by comparing the dispersion relations in the $45$\,nm-thick plain film and the $10$\,nm-thick film slice under the $35$\,nm-deep nanogroove, as depicted in \textcolor{blue}{Figure} \ref{fExperiment}(a). These dispersion relations are plotted by the dashed and dotted lines in \textcolor{blue}{Figure} \ref{fExperiment}(b), respectively. The phase shift acquired by spin waves as they pass through the groove region can be calculated as $\Delta \Theta = b(k_2 - k_1) = b \Delta k_{12}$, where $k_1$ and $k_2$ are the spin-wave wave vectors in the respective waveguides, \textcolor{blue}{Figure} \ref{fExperiment}(b). At $20$\,GHz, which is an exemplary frequency for the spin-wave propagation in the saturated state, we come with $\Delta\Theta = 43^\circ$ that is by a factor of four smaller than the experimentally measured value. Hence, this estimate is strongly indicative of additional effects involved in the phase-shift mechanism.

\begin{figure}
    \centering
    \includegraphics[width=1\linewidth]{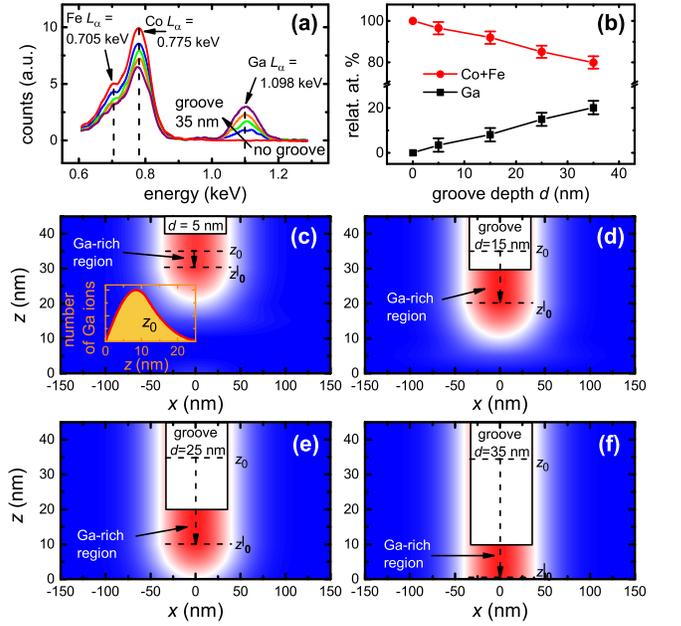}
    \caption{\textbf{Material composition in the Co-Fe waveguides}.
    (a) Energy-dispersive x-ray spectra for the as-deposited waveguide in comparison with the waveguides with nanogrooves. The data were normalized in such a way that the sum of the content of [Ga] and [Co+Fe] in the sample is 100 at.\%. The peaks of Co, Fe, and Ga are shown by vertical dashed lines.
    (b) Material composition in the waveguides as a function of the groove depth.
    (c)-(f) Distributions of 30\,kEV Ga ions stopped in the Co-Fe waveguides under the nanogroove obtained from Monte Carlo simulations. The number of Ga ions is increasing from zero (away from the groove, blue) to a maximum value (under the groove, red). Vertical arrows indicate how the peak in the distribution of the Ga ions stopped in the waveguide [inset in (c)] has moved during the FIB milling of the nanogroove from its initial position $z_0$ to the final position $z_0^\prime$. The origin and head of the dashed arrows correspond to the start and the end of the FIB milling process, respectively.}
    \label{fEDX}
\end{figure}

To elucidate the origin of the phase inversion in our system, we next discuss the results of micromagnetic simulations presented in \textcolor{blue}{Figure} \ref{fExperiment}(d) and detailed in \textcolor{red}{Supporting Information}. Namely, from the calculated dependence $\Delta \Theta(d)$ it follows that if one considers the nanogroove solely as a geometrical thinning of the waveguide, then even the deepest nanogroove is not expected to induce a spin-wave phase shift exceeding $70^\circ$. Therefore, incorporation of Ga ions under the milled groove has been considered very likely and a microstructural characterization of the Co-Fe waveguides in the vicinity of the FIB-milled groove has been performed. The results of material composition analysis by energy-dispersive x-ray (EDX) spectroscopy are summarized in \textcolor{blue}{Figure} \ref{fEDX}. The EDX data in \textcolor{blue}{Figure} \ref{fEDX}(a) and (b) indicate that the Ga content in the $200\times200$\,nm$^2$ region encaging the groove increases from about $5$ at.\% to $20$ at.\% as the groove depth reaches $35$\,nm. Given that the waveguide section probed by EDX is a factor of $1.65$ larger than the groove width, the more accurate contents of Ga are expected to be between $8$ at.\% and $35$ at.\% in the layer under the groove. Monte-Carlo simulations of the distribution of Ga ions stopped in the waveguides presented in \textcolor{blue}{Figure} \ref{fEDX}(c)--(f) suggest that the Ga ions stop within an about $10$\,nm-thick layer under the groove and, to some smaller extend, within a $5$\,nm-thick layer at the nanogroove walls. Accordingly, the incorporation of Ga is expected to lead to a local reduction of the saturation magnetization $M_\mathrm{s}$ under the nanogroove.

Calculated spin-wave phase shifts $\Delta \Theta(d)$ at $20$\,GHz for the assumed reduced magnetization in a $10$\,nm-thick slice of the Co-Fe film just below the $35$\,nm-deep groove are shown in \textcolor{blue}{Figure} \ref{fExperiment}(d). In both cases $\Delta \Theta$ reaches about $300^\circ$ at a groove depth of $35$\,nm, pointing to the decisive effect of the reduced-$M_\textrm{s}$ layer on the phase shift in the waveguide. Furthermore, one immediately notes three important features of the $\Delta \Theta(d)$ curves emerging when a reduced $M_\textrm{s}$ is taken into account. Firstly, for films without a groove ($d=0$), the reduced $M_\textrm{s}$ alone leads to a phase shift of about $15^\circ$ and $30^\circ$, respectively. Secondly, $\Delta \Theta$ increases slowly up to some critical thickness, $d_\mathrm{c}$, which amounts to about $23$\,nm and $26$\,nm for the $10$\,nm-thick layers with $0.5M_\textrm{s}$ and $0.65M_\textrm{s}$, respectively. Thirdly, and most importantly, at $d_\mathrm{c}$ the phase abruptly jumps by about $180^\circ$, i.e. a complete phase inversion takes place. In all, the significant increase of $\Delta\Theta$ evidences that the reduced-$M_\mathrm{s}$ mechanism is dominating the geometric mechanism. We attribute the almost linear increase of the experimental dependence $\Delta\Theta(d)$ in the inset in \textcolor{blue}{Figure} \ref{fExperiment}(c) to an averaging of the simulated curves $\Delta\Theta(d)$ in \textcolor{blue}{Figure} \ref{fExperiment}(d) for different $M_{\mathrm{s}}$ values, due to a gradual magnetization variation as a function of the $z$-coordinate under the nanogroove. We believe that some contribution to smearing of the abrupt phase jump may also be caused by the rounded corners of the nanogroove (i.e. a groove width gradient) so that the critical depth $d_\mathrm{c}$ is experimentally smeared.

Finally, we return to \textcolor{blue}{Figure} \ref{fExperiment}(b) and consider the dispersion relation for the film with reduced both, thickness and magnetization, in comparison with the plain film. The dispersion relation of the former is shifted towards lower frequencies due to the reduced magnetization and it is flattened owing to the smaller thickness. At a fixed frequency above the ferromagnetic resonance frequency in the homogeneous part, the large difference in the wavenumbers between the dispersions results in the phase accumulation $\Delta k_{13}$. This elucidates the large phase shifts observed experimentally, as for the waveguide with the $35$\,nm-deep nanogroove we obtain $\Delta\Theta\approx245^\circ$. This phase shift is only slightly larger than in the experiment, where $\Delta \Theta (20\,\mathrm{GHz}) \approx 215^\circ$. Hence, simulations nicely agree with experiment. The small discrepancy between the calculated and measured values is attributed to the rounded corners of the nanogroove and a more complex distribution of Ga ions stopped in the waveguide.

In all, our findings have several important implications for both, the domains of magnonics and focused particle beam-induced processing. The main experimental finding --- the complete phase inversion in the spin-wave propagation through a single nanogroove --- is a robust effect which systematically evolves in all grooved waveguides as a function of the biasing magnetic field and the groove depth. The key element of the phase shifter is a nanogroove, that is a typical defect fabricated by focused ion beam milling that is widely used in many areas of condensed matter physics and materials science \cite{Utk08vst}. While FIB-milled defects have previously been shown to guide magnetization reversal in circular Co-FEBID disks through metastable magnetic configurations \cite{Lar14apl}, here FIB-milled nanosized defects are used to affect the dynamic response of microscopic FEBID conduits. While the lateral resolution of FEBID rivals that of advanced electron beam lithography, FEBID is in addition capable of creating binary and ternary alloy systems as well as complex three-dimensional nano-architectures \cite{Dob15bjn,Por15nan,Hut18mee}. Advantages of this mask-less nanofabrication technique are already appreciated in materials science and magnetism, and here we have unveiled its capabilities for nano-magnonics.

We also stress that the microscopic mechanism of the spin-wave phase shift in our system is physically different from that in Ref. \cite{Bau18apl} and our nanogrooves are one-to-two orders of magnitudes smaller than the magnetic defects used in Ref. \cite{Bau18apl}. Given the resolution capability of FIB milling, further downscaling of nanogrooves down to a width of $30$\,nm \cite{Dob11snm} can be anticipated. In addition, FIB milling allows one to pre-define the steepness of the groove slopes differently \cite{Dob17nsr} and, thereby, to introduce a non-reciprocity into the system \cite{Mru17prb}. From the broader perspective of nano-magnonics, the proposed phase shifters can be realized in continuous microscopic waveguides which are state-of-the-art elements of circuitry. Furthermore, due to the flexibility of the shape of FIB-milled grooves, they can be shaped, i.g. as arcs and even depth gradients can be introduced to tailor the spin-wave front. Finally, given the large magnetostriction in body-centered Fe-based solid solutions \cite{Jen10jap,Lis14apl}, further prospects for the interplay of spin waves with surface acoustic waves can be anticipated \cite{Gra17prb}.

\section{Conclusion}
In summary, we have successfully realized a tunable spin-wave phase shifter upon a single nanogroove milled by focused ion beam in a Co-Fe microsized magnonic waveguide. By increasing the groove depth and the in-plane biasing magnetic field we have continuously tuned the spin-wave phase and experimentally evidenced a complete phase inversion. The microscopic mechanism of the spin-wave phase inversion is based on the combination of geometrical and the reduced-magnetization mechanisms in the Ga-rich slice of the waveguide just below the nanogroove. The proposed phase shifter can be on-chip integrated with spin-wave logic gates and other magnonic devices. Exploiting the widely available focused ion beam milling as a mask-less nanofabrication tool, our findings have important implications for phase-tuning of spin waves in magnonic circuits and the generation of spin-wave beams by phased-array elements in spin-wave nano-optics.

\section{Experimental section}
\label{sExp}

\textbf{Fabrication of coplanar antennas.}
Phase shifters were fabricated on top of $55$\,nm-thick antennas milled by focused ion beam (FIB) from an Au film in a high-resolution scanning electron microscope (SEM: FEI Nova NanoLab 600). The SEM was equipped with a Schottky electron emitter and operated at a base pressure of about $2\times10^{-7}$\,mbar. The Au film was prepared by dc magnetron sputtering onto Si/SiO$_2$ substrate with a pre-sputtered 5\,nm-thick Cr buffer layer. The thickness of the SiO$_2$ layer was 200\,nm. In the sputtering process the substrate temperature was $T = 22^\circ$C, the growth rate was $0.055$\,nm/s and $0.25$\,nm/s, and the Ar pressure was $2\times10^{-3}$\,mbar and $7\times10^{-3}$\,mbar for the Cr an Au layers, respectively. The center-to-center distance between the signal lines of the CPWs amounted to $2.5\,\mu$m. The width of the signal and ground lines of the CPWs was $w =200$\,nm while the gap between them amounted to $s =120$\,nm, to match the $50\,\Omega$ impedance of the transmission line. A $10$\,nm-thick isolating TiO$_2$ layer was deposited on top of the CPWs by FEBID employing the titanium(IV)-isopropoxide precursor, prior to the deposition of the Co-Fe magnonic waveguides.

\textbf{Fabrication of magnonic waveguides.}
The Co-Fe waveguides were fabricated by FEBID employing the precursor HCo$_3$Fe(CO)$_{12}$. The precursors were injected in the SEM through a capillary with an inner diameter of $0.5$\,mm. The distance capillary-surface was about $100\,\mu$m and the tilting angle of the injector was $50^\circ$. The crucible temperature of the gas injection system for the Co-Fe process was set to $65^\circ$. The waveguides are $(45\pm1)$\,nm thick and ($1000\pm3$)\,nm wide. They were grown in the high-resolution deposition mode with $5$\,keV beam energy, $1.6$\,nA beam current, $20$\,nm pitch, and $1\,\mu$s dwell time. The same beam parameters were used for the deposition of the protective TiO$_2$ layer. The growth rate of the waveguides was about $2$\,nm/min. The elemental composition in the waveguides is about 82 at\% [Co+Fe], 10 at\% [O] and 8 at\% [C], as inferred by energy-dispersive x-ray spectroscopy. Previous microstructural investigations of the Co-Fe deposits by transmission electron microscopy revealed that the waveguides consist of a dominating bcc Co$_3$Fe phase mixed with a minor amount of FeCo$_2$O$_4$ spinel oxide phase with nanograins of about $5$\,nm diameter \cite{Por15nan}. The waveguides are characterized by a saturation magnetization $M_\textrm{s}$ of $1.4\times10^6$\,A/m, as deduced from reference ferromagnetic resonance measurements.

\textbf{Fabrication of nanogrooves.}
One waveguide was left as-deposited (plain), while a nanogroove was milled by FIB in the middle of each of four other waveguides. The FIB milling of the nanogrooves was done with a Ga beam at a beam voltage of $30$\,kV and a beam current of $10$\,pA. For high-resolution characterization of the fabricated nanogrooves, atomic force microscopy (AFM) under ambient conditions in non-contact, dynamic force mode was used. The cantilever tip was shaped like a polygon-based pyramid, with a tip radius of less than $7$\,nm (Nanosensors PPP-NCLR), so that convolution effects due to the finite tip radius can be neglected. The grooves have not a rectangular, but slightly smeared profile, so that we defined the width $b$ of the grooves at their half depth. While the width $b = (120\pm2)$\,nm was kept constant in all grooved waveguides, we varied the groove depth $d$ between $5$\,nm and $35$\,nm in steps of 10\,nm with an accuracy of $1$\,nm. An rms surface roughness of less than $0.2$\,nm over a scan field of $1\times1\,\mu$m$^2$ was inferred from the AFM inspection of the waveguides.

\section*{Acknowledgements}
Research leading to these results received funding from the European Union's Horizon 2020 research and innovation program under Marie Sklodowska-Curie Grant Agreement No. 644348 (MagIC). Furthermore, this article is based upon work from Action CA16218 (NANOCOHYBRI) of the European Cooperation in Science and Technology (COST). The Portuguese team acknowledges the Network
of Extreme Conditions Laboratories-NECL and Portuguese Foundation of Science and Technology (FCT) support through the projects NORTE-01-0145-FEDER-022096, MIT-EXPL/IRA/0012/2017,  POCI-0145-FEDER-030085 (NOVAMAG), EXPL/IF/00541/2015 (S.A.B.), EXPL/IF/01191/2013 (D.N.), and EXPL/IF/00981/2013 (G.N.K).

\section*{Supporting Information}

\begin{figure}[t!]
    \centering
    \includegraphics[width=1\linewidth]{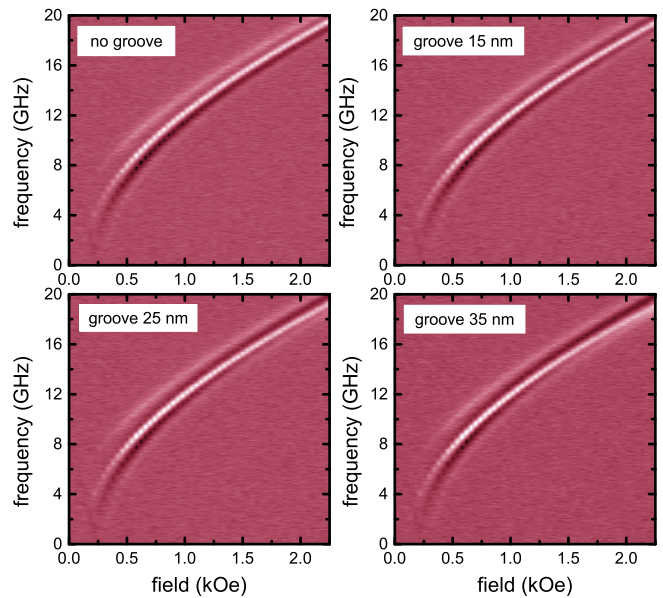}
    \caption{\textbf{Transmission of spin waves through the waveguides monitored by all-electrical spin-wave spectroscopy}. Magnitude of the forward transmission coefficient $S_{21}(f)$ is shown for increasing magnetic field. Note the gradually reverting contrast (related to the spin-wave phase shift) from sample to sample with an increase of the groove depth. The interchange of the dark and bright branches is best seen at higher magnetic fields.}
    \label{fFMRs}
\end{figure}

\textbf{All-electrical spin-wave spectroscopy.}
A vector network analyzer (VNA) was used to apply a radiofrequency signal with a power of $-10$\,dBm to CPW1 and detect the spin wave transmission at CPW2, Supplementary Figure \ref{fFMRs}, by all-electrical spin-wave spectroscopy \cite{Neu10prl,Tac12prb}. Two picoprobes (GGB Industries 40A-GSG-150-S), positioned by fine-tuning of translational microscrews under a light microscope, were used for connecting the samples to the VNA. The transmission of spin waves was analyzed in terms of the absolute value of the forward transmission coefficient $S_{21}(H,f)$ recorded as a function of frequency and biasing magnetic field value. Here, $S_{21}$ is a measure of the microwave power received at CPW2 in comparison with that applied at CPW1. The magnitude of the forward scattering coefficient $S_{21}$ was recorded as a function of frequency $f$ with the magnetic field applied in the $y$-direction, i.e. in the Damon-Eshbach geometry. To extract the spin-wave contribution, the reference spectrum $S_{21}(f)$ at $H = 0$, where the spin-wave excitation is negligible, was subtracted from the measured data.

From the most pronounced neighboring extrema in all data sets in Supplementary Figure \ref{fFMRs} relative signal strength $\Delta_{P1-P2}$ and the frequency difference $\delta f = f_{P2}-f_{P1}$ between them have been deduced. To quantify the amplitude change and the frequency and phase shifts in the waveguides with nanogrooves, we introduce $\Delta f = f^{\mathrm{groove}}_{P2} -f^{\mathrm{ref}}_{P2}$ evaluated at different $H$ for the $P2$ peaks for the grooved and reference samples. The resulting dependences $\Delta f(H)$ for the four samples are shown in Supplementary Figure \ref{fExperiments}(a). The frequency shift is negative and it increases with $H$. The deeper the nanogroove, the larger the frequency shift which attains $\Delta f \approx -0.5$\,GHz at $H = 2.25$\,kOe. The dependence of the frequency shift on the nanogroove depth at $H = 2$\,kOe is plotted in the inset of Supplementary Figure \ref{fExperiments}(a).
For completeness, the magnetic field dependence of the phase shift $\Delta\Theta (H)$ from Fig. 3(c) of the manuscript is shown in Fig. \ref{fExperiments}(b).

Upon the transmission of spin waves through the nanogrooves, the relative signal strength $\eta = \Delta^{\mathrm{groove}}_{P2-P1}/\Delta^{\mathrm{ref}}_{P2-P1}$ decreases almost linearly with both, increasing $H$ and $d$, Supplementary Figure \ref{fExperiments}(c). Introducing the transmission attenuation parameter $A$ defined as the ratio of the magnitude of $S_{21}(H)$ for a grooved waveguide to its magnitude in the plain one we reveal that even for the 35\,nm-deep nanogroove the relative transmission decrease does not exceed $40\%$, see Supplementary Figure \ref{fExperiments}(d). The dependence of $A$ on $d$ in the inset of Supplementary Figure \ref{fExperiments}(d) suggests that a $50\%$ reduction of the waveguide thickness results in only a $20\%$ reduction of $A$ while a further increase of the groove depth leads to a more rapid transmission attenuation. The insets in Supplementary Figure \ref{fExperiments} display the dependence of the transmission characteristics on the groove depth at a biasing magnetic field of $2$\,kOe corresponding to an envelope's central frequency of about $18.6$\,GHz. As it follows from these dependences, all transmission characteristics exhibit a systematic dependence on $d$, attesting to that the groove depth can be considered as an important nanostructure parameter for tuning of the spin-wave phase and amplitude.
\begin{figure}[t!]
    \centering
    \includegraphics[width=0.95\linewidth]{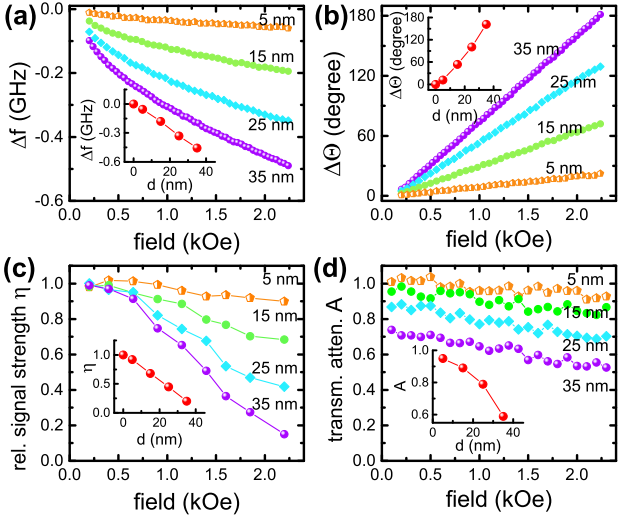}
    \caption{\textbf{Tuning spin-wave transmission parameters by magnetic field and groove depth variation}. (a) Frequency difference $\Delta f = f_{P2}^{\mathrm{groove}} - f_{P2}^{\mathrm{ref}}$ between the main peaks for the grooved and reference waveguides. Inset: Dependence of $\Delta f$ on the depth of the nanogroove at $H = 2$\,kOe. (b) Phase shift $\Delta\Theta$ as a function of $H$ and the groove depth $d$ at $H = 2$\,kOe (inset). (c) Relative signal strength $\eta = \Delta^{\mathrm{groove}}_{P2-P1}/\Delta^{\mathrm{ref}}_{P2-P1}$ as a function of $H$  and $d$ at $H = 2$\,kOe (inset). (d) Transmission attenuation parameter $A = S_{21}^{\mathrm{groove}}/S_{21}^{\mathrm{ref}}$ as a function of $H$ and $d$ at $H = 2$\,kOe (inset).}
    \label{fExperiments}
\end{figure}

\textbf{Material composition in the waveguides.}
The gallium ion beam is known to cause amorphization, implantation, and vacancy generation in the near-surface area of the processed region~\cite{Utk08vst}. In particular, calculations using Monte Carlo simulations~\cite{Zie85boo} indicate that an implantation of Ga ions occurs chiefly in a depth of up to approximately $10$\,nm of the Co-Fe waveguide. The stopping of the Ga ions in the Co-Fe waveguide was simulated with the help of the SRIM software. An inspection using energy-dispersive x-ray (EDX) spectroscopy was performed in the SEM right after the fabrication of grooves, without exposure of the samples to air. The EDX test area was $200\times200$\,nm$^2$, and the EDX parameters were $5$\,kV and $1.6$~nA. Here the beam energy determines the effective thickness of the layer being analyzed, which is approximately 45\,nm. The penetration of the electrons into the film was calculated by the simulation program Casino. This corresponds to approximately $90\%$ of the electron beam energy dissipated in the film. The material composition was calculated taking ZAF (atomic number, absorbtion, and fluorescence) and background corrections into account. The software used to analyze the material composition in the film was EDAX's Genesis Spectrum v.~5.11. The statistical error in elemental composition is $1.5\%$. The spatial distribution of the number of stopped ions has a maximum at a depth of $10$\,nm from the groove bottom. The EDX data were normalized such that the sum of the atomic percentage of Co, Fe, and Ga amounts to $100\%$.

\textbf{Micromagnetic simulations.}
Micromagnetic simulations were performed using MuMax 3 \cite{Van14aip}. First, the static magnetization configuration was stabilized for a given magnetic field. Next, simulations of the spin-wave dynamics were performed. In simulations, a saturation magnetization of $M_s = 1.4 \times 10^6$\,A/m and an exchange constant of $A = 1.4 \times 10^{-11}$\,J/m were used. From the first part of simulations, a hysteresis loop was extracted, revealing a $M(H)$ dependence typical for the magnetization reversal process along the hard axis. This points to the influence of the finite width of the magnonic waveguide. Below a magnetic field value of about $1$\,kOe, the remagnetization process sets on. At $H = 0$ the magnetization is oriented along the stripe axis, where excitation of spin waves is ineffective, in agreement with the experimental results. To elucidate the microscopic mechanism responsible for the large phase shifts observed experimentally, we performed spin-wave dynamic simulations with periodic boundary conditions along the $y$ axis at $H = 2$\,kOe, i.e. in the saturated state in which static demagnetizing effects are absent. The spin waves were excited by a continuous harmonic signal of small amplitude at $20$\,GHz at a distance of $1$\,$\mu$m from the nanogroove. The simulated time-dependent signal was collected through the whole structure. The phase shift was extracted in reference to the data obtained for the plain waveguide. The nanogroove was modeled as a rectangular groove with a width of $120$\,nm and the groove depth varied between $2.5$\,nm and $35$\,nm in steps of $2.5$\,nm.

\clearpage


%

\end{document}